\documentclass[11pt]{article}
\pdfoutput=1
\usepackage{tikz}
\usepackage{amssymb,amsmath,amsfonts}
\usepackage{latexsym}
\usepackage[totalwidth=17cm,totalheight=23.8cm]{geometry}

\usepackage{feynmp}
\usepackage[force]{feynmp-auto}

\newcommand{\C}{{\mathbb C}}
\newcommand{\R}{{\mathbb R}}

\newcommand{\im}{{\rm i }}

\newcommand{\ket}[2]{ {\langle #1\,#2\rangle} }
\newcommand{\bra}[2]{ {[#1\,#2]}}

\newcommand\be{\begin{eqnarray}}
\newcommand\ee{\end{eqnarray}}

\begin{document}

 \unitlength = 1mm

 \begin{fmffile}{uniq}

\title{GR uniqueness and deformations}
\author{Kirill Krasnov \\ \it{School of Mathematical Sciences, University of Nottingham}\\ \it{University Park, Nottingham, NG7 2RD, UK}}

\date{v2: July 2015}
\maketitle

\begin{abstract} In the metric formulation gravitons are described with the parity symmetric $S_+^2\otimes S_-^2$ representation of Lorentz group. General Relativity is then the unique theory of interacting gravitons with second order field equations. We show that if a chiral $S_+^3\otimes S_-$ representation is used instead, the uniqueness is lost, and there is an infinite-parametric family of theories of interacting gravitons with second order field equations. We use the language of graviton scattering amplitudes, and show how the uniqueness of GR is avoided using simple dimensional analysis. The resulting distinct from GR gravity theories are all parity asymmetric, but share the GR MHV amplitudes. They have new all same helicity graviton scattering amplitudes at every graviton order. The amplitudes with at least one graviton of opposite helicity continue to be determinable by the BCFW recursion. 
\end{abstract}

\section{Introduction} 

The statement of GR uniqueness is that General Relativity is the only theory of interacting massless spin two particles without higher derivatives in field equations. Its existing proofs make a seemingly innocuous assumption that the graviton is described by a symmetric rank two tensor. This assumption, together with the requirement that there are not higher than second derivatives in the field equations, allows one to either completely fix the form of the Lagrangian (modulo field redefinitions and total divergences), or fix the form of the simplest tree-level graviton scattering amplitudes, with everything else being determined by factorisation, see below.  

The purpose of this paper is to show that the above GR uniqueness statement rests crucially on the assumption that the graviton is described by the metric. Thus, we show that, if one allows oneself to use a different representation of the Lorentz group to describe gravitons, there is not a single, but {\it an infinite-parameter family of theories of interacting massless spin two particles without higher derivatives in field equations.} 

There is also an important caveat to the above statement. By General Relativity one usually understands a dynamical theory of real-valued Lorentzian signature metrics. This theory can be studied perturbatively around the Minkowski space-time background, and gives rise to a unitary S-matrix. However, it is also of interest to consider GR as a theory of Riemannian or split signature metrics. More generally, one can complexify the manifold in question, and consider GR as a theory of complex-valued metrics depending holomorphically on all the coordinates. Then the statement of GR uniqueness continues to apply, see below. Our claim of existence of interacting theories of gravitons distinct from GR (once a different representation of Lorentz is used) applies in the complexified setting. At the moment we do not know whether there exist reality conditions that can be imposed to convert the distinct from GR gravity theories into unitary interacting theories. 

The distinct from GR theories of interacting gravitons (in 4 space-time dimensions) have been around for quite some time, with the first reference mentioning their possibility being \cite{Capovilla:1989ac}. Bengtsson \cite{Bengtsson:1990qg} was the first to point out that there is an infinite-parameter family of such theories, dubbing them "neighbours of GR". They were later rediscovered \cite{Krasnov:2006du} and extensively studied by the present author. Hamiltonian analysis of these theories \cite{Krasnov:2008zz} shows that they continue to propagate just two polarisations of the graviton. Thus, they can be continuously deformed back to GR by setting to zero the new coupling constants. In this sense these theories can be referred to as deformations of General Relativity. 

Unfortunately, the existing literature on this subject is not easy to penetrate without some background in "self-dual" variables. In particular, there is no simple description of what these theories are in the metric language. This is one explanation of why the existence of these distinct from GR theories of interacting gravitons has been mostly ignored by the theoretical physics community. 

The purpose of this article is to explain how the uniqueness of GR is avoided in as simple terms as possible. To this end we shall employ the language used by one of the most convincing proofs of GR uniqueness --- the language of spinor helicity and scattering amplitudes. This language can be used without much change even in the adopted here complexified setting. More or less, one simple allows all objects to be complex-valued. 

Thus, there is by now a widely known argument fixing the form of the 3-graviton amplitudes from their transformation properties under the little group. Simple dimensional analysis then shows that only the $++-$ and $--+$ amplitudes can come from a vertex without higher derivatives, while the $+++$ and $---$ amplitudes must be zero if no higher derivatives is allowed. We will show that the choice of a different representation of the Lorentz group to describe gravitons changes the dimensional count, and that it becomes possible to have the $+++$ amplitude (or $---$ but not both) without higher derivatives in field equations. Further, it turns out to be possible to have new all plus amplitudes (and thus new couplings) at every graviton order, the result being an infinite-parametric family of theories of interacting gravitons without higher derivatives in field equations. 

Our arguments are as simple and as general as those used in the amplitude-based proof of GR uniqueness. We thus hope that the arguments below establish the existence of distinct from GR parity-violating theories of interacting massless spin 2 particles beyond any doubt.

The deformations of GR that we describe have non-trivial mostly plus amplitudes, while the mostly minus amplitudes continue to be zero as in GR, see below for our conventions on the graviton helicity. Therefore, in the Lorentzian signature, the amplitudes assemble into an S-matrix that is not unitary, at least not unitary with respect to the inner product one would normally use in this context. So, at the moment the only available interpretation of these theories is as giving deformations of complexified General Relativity while keeping both the number of propagating degrees of freedom and the order of field equations intact. Interpreted this way, the problem of finding a Lorentzian signature physical interpretation reduces to the problem of determining appropriate reality conditions for the fields. This is an open problem, which is the second reason why the theories we describe are not as well known as they perhaps deserve to be. 

We start by reviewing in Section \ref{sec:GR} the amplitude-based proof of GR uniqueness. This Section also sets our notations. Section \ref{sec:conn} explains how gravitons can be described by a different representation of the Lorentz group, and then performs the dimensional analysis that leads to the conclusion of possibility of having the $+++$ amplitudes. In Section \ref{sec:unitarity} we show how more complicated graviton amplitudes may be constructed from factorisation, similarly to what happens in the case of GR. Finally, in Section \ref{sec:BCFW} we show how the BCFW recursion may still be used to determine large classes of amplitudes. We conclude with a discussion.

To avoid confusion, we remark that everywhere in this paper the word "graviton" stands for "massless spin 2 particle". As we explain below, the later can be described in several different ways, only one of which uses the familiar metric perturbations. 

\section{Amplitude-based proof of GR uniqueness}
\label{sec:GR}

\subsection{Spinors and helicity}

The material described here is standard, see e.g. subsection 1.2 of \cite{ArkaniHamed:2008gz} for a nice discussion. We first describe notions appropriate for Minkowski signature, and then turn to the complexified setting. 

A (future oriented) null vector in 4 space-time dimensions can be represented as a product of two spinors $k_\mu\equiv k_{AA'} = k_A k_{A'}$. Here $A,A'$ are the unprimed and primed spinor indices. The subgroup of the (Lorentzian signature) Lorentz group fixing $k_\mu$ (the little group) is isomorphic to the group of isometries of the Euclidean 2-plane. This contains ${\rm SO}(2)\sim {\rm U}(1)$ that acts on spinors $k_A, k_{A'}$ by multiplying $k_A \to e^{\im\theta} k_A, k_{A'}\to e^{-\im\theta} k_{A'}$, keeping them complex conjugates of each other. 

States of given helicity are those transforming in a particular way under the transformations from ${\rm U}(1)$. In the metric formalism one describes gravitons using symmetric rank two tensors $h_{\mu\nu}$. In the spinor language symmetric rank two tensors translate into $h_{ABA'B'}$, which is symmetric in pairs $AB$ and $A'B'$, as well as the trace part. The physical gravitons are described by $h_{ABA'B'}$. As is well known, using the gauge freedom of gravity (diffeomorphisms) the two physical polarisations of the graviton can be described by the following helicity spinors
\be\label{helicity-h}
\epsilon^-_{ABA'B'} = \frac{q_A q_B k_{A'} k_{B'}}{\ket{q}{k}^2}, \qquad
\epsilon^+_{ABA'B'} = \frac{k_A k_B q_{A'} q_{B'}}{\bra{k}{q}^2}.
\ee
where $q_A, q_{A'}$ are two reference spinors (usually not chosen to be related by complex conjugation even in Lorentzian signature), and $\ket{\lambda}{\mu} := \lambda^A \mu_A, \bra{\lambda}{\mu}:= \lambda_{A'} \mu^{A'}$ are the spinor contractions. The spinors $k_A, k_{A'}$ in each helicity spinor are the momentum spinors of the corresponding particle. Note that under the little group $\epsilon^\pm \to e^{\pm 4\im\theta} \epsilon^\pm$, which corresponds to helicities $\pm 2$.

\subsection{Complexified setting}

In the discussion of scattering amplitudes one usually has in mind that everything is happening on the Minkowski space-time background. However, as we now review, many statements continue to be true in the complexified setting as well.  

One can treat metrics of Lorentzian, Riemannian and split signatures in a unified way by passing to the complexified GR. To this end, one makes all the manifold coordinates complex (thus considering a complexified manifold), and considers metrics that are holomorphic functions of the coordinates. Then appropriate real sections give metrics of Lorentzian, Riemannian or split signatures. 

One can also consider complexification of the scattering theory. Thus, one uses the flat metric $\eta_{\mu\nu}={\rm diag}(1,1,1,1)$ as the background, but allows the coordinates, as well as the metric perturbation to become complex. One can then formally solve the linearised field equations in terms of "gravitons" characterised by complex momenta $k_\mu$ with $k^\mu k_\mu=0$. Formally, there are still two possible types of linearly independent solutions of the second order field equations -- the "positive" and "negative" frequency ones. For each of these two types, modulo gauge, there are still just two graviton polarisations. 

Since in the complexified setting the metric perturbation is no longer required to be real, there is no relation between the amplitudes of the "positive" and "negative" frequency waves. Thus, there is neither the notion of creation and annihilation operators, nor LSZ reduction. However, one can still define the notion of the correlation functions of a product of metric operators (e.g. using the functional integral). Such a correlation function can then be Fourier transformed (at least formally, for some suitable choice of the integration countour). One can then define the "S-matrix" as being composed of the residues of the Fourier transformed correlation functions as the momenta go null $k^2\to 0$. 

Throughout the paper we will continue to talk about the above defined object as an "S-matrix" even though it is a completely holomorphic object, constructed without ever using the operation of complex conjugation. As unitarity relates the S-matrix to its complex conjugate, the S-matrix of our complexified setting cannot be a unitary object (even though it may give rise to a unitary one once it is appropriately restricted by the reality conditions). At least at tree level, the S-matrix of the complexified setting is a convenient object, as it encodes the perturbative solutions of the complexified field equations.  

The complexified setting is also convenient because one no longer has to worry about reality of the momentum vectors. Since the manifold coordinates became complex, so are the momenta. Once the momentum is complexified, the two types of (complexified) Lorentz group spinors are no longer related to each other. Different signature choices are then different real slices. For example, in the split signature the spinors $k_A, k_{A'}$ are taken to be two independent real spinors, with the split signature Lorentz group being ${\rm SL}(2,\R)\times {\rm SL}(2,\R)$. To get the Euclidean signature one works with complex valued independent spinors $k_A, k_{A'}$, with a certain operation of complex conjugation that acts on each of the two spaces $k_A$ and $k_{A'}$ without mixing them. 

In the complexified setting the little group ${\rm U}(1)$ gets complexified into $\C_*$, the group of multiplication by a complex number different from zero, with the following action $k_A \to t k_A, k_{A'}\to t^{-1} k_{A'}$. In the split signature case the parameter $t$ must be real. 

\subsection{3-graviton amplitudes}

The 3-graviton amplitudes are uniquely fixed by their transformation properties under the little group. Indeed, given a triple $k_{1,2,3}$ of null $k_i^2=0$ vectors satisfying momentum conservation $k_1+k_2+k_3=0$, it is easy to show that there are just two possible ways in which such a configuration can be represented by spinors: (i) either all unprimed spinors are multiples of each other, so that all angle bracket contractions vanish or (ii) all primed spinors are multiples of each other so that all square bracket contractions vanish. 

Let us assume that we are dealing with a configuration in which all square bracket contractions vanish. Then a 3-graviton amplitude, being Lorentz invariant, must be a function of the angle bracket contractions $\ket{1}{2}, \ket{1}{3}, \ket{2}{3}$, where we use the notation $(k_1)_{AA'} = (k_1)_A (k_1)_{A'} \equiv 1_A 1_{A'}$, etc. Then the fact that the amplitude must transform correctly under the individual little group transformations $1_A \to t_1 1_A, 2_A\to t_2 2_A, 3_A \to t_3 3_A$ implies that the amplitudes are multiples of the following expressions
\be
{\cal M}^{-++} \sim \frac{\ket{2}{3}^6}{\ket{1}{2}^2 \ket{1}{3}^2}, \qquad
{\cal M}^{+++} \sim \ket{1}{2}^2\ket{1}{3}^2\ket{2}{3}^2. 
\ee
It is also possible to have the other two helicity configurations $--+$ and $---$ constructed from the angle brackets, just by taking the inverses of what appears on the right-hand-sides of the above expressions. But we shall soon see that only the above choice is physically sensible.

Similarly, if one deals with a configuration in which only the square brackets are non-vanishing then the following amplitudes result
\be
{\cal M}^{+--} \sim \frac{\bra{2}{3}^6}{\bra{1}{2}^2 \bra{1}{3}^2}, \qquad
{\cal M}^{---} \sim \bra{1}{2}^2\bra{1}{3}^2\bra{2}{3}^2. 
\ee
The other two helicity configurations can also be obtained by taking the inverses, but this is not a physically interesting solution, as we shall see in the next subsection.

\subsection{Dimensional analysis}

The 3-graviton amplitudes must have the mass dimension 1, while the above expressions have the mass dimension either 2 or 6, as each pair of brackets carries the mass dimension one. This means that the correct amplitudes are given by (suppressing factors of the imaginary unit)
\be\label{-+}
{\cal M}^{-++} = \frac{1}{M_p} \frac{\ket{2}{3}^6}{\ket{1}{2}^2 \ket{1}{3}^2}, \qquad
{\cal M}^{+++} = \frac{\alpha}{M_p^5} \ket{1}{2}^2\ket{1}{3}^2\ket{2}{3}^2. 
\ee
Here $M_p$ is some scale, with the first of these amplitudes defining this scale, and $\alpha$ is some dimensionless coupling constant, that does not have to be order unity. To be precise, what arises in the case of GR in front of the first of these amplitudes is $\sqrt{2}/M_p$, with $M_p$ being the usual Planck mass $M_p^{-2} = 16\pi G$, but such details are not going to matter in what follows. 

Let us now see what kind of Lagrangian such amplitudes can come from. It is clear that to obtain the first of these amplitudes one must have a cubic interaction with a factor of $1/M_p$ in front. Given that the sought cubic term in the Lagrangian must be of mass dimension 4 (we are in 4 space-time dimensions), and that our helicity spinors (\ref{helicity-h}) do not contain any dimensionful parameters, we see that the interaction vertex in question must be of the form
\be
{\cal L}\sim \frac{1}{M_p} h (\partial h)^2,
\ee
where $\partial$ is the partial derivative. Of course there may be some complicated tensorial structure here, but this is of no importance for us. What is important is that the $-++$ amplitude in (\ref{-+}) can come from an interaction term in the Lagrangian with just two derivatives. Such a Lagrangian would lead to second order in derivatives field equations, which is acceptable of a physical theory. So, we conclude that the first of the amplitudes in (\ref{-+}) is acceptable. Of course, it is this amplitude that arises in GR.

Let us now analyse the second of the amplitudes in (\ref{-+}). The same type of dimension counting argument shows that it can only come from a cubic interaction term with as many as 6 derivatives
\be\label{W3}
{\cal L} \sim \frac{\alpha}{M_p^5} (\partial^2 h)^3.
\ee
Such an interaction term would clearly lead to field equations with higher derivatives. This is known to lead to instabilities, and so we must discard the $+++$ amplitude in (\ref{-+}) as being impossible in a theory of interacting gravitons with not higher than second order field equations. Of course, such an amplitude may be non-zero in some healthy theory that contains (\ref{W3}) as a part of its low-energy effective metric Lagrangian. But (\ref{W3}) leading to higher derivatives is not possible in a {\it fundamental} theory, i.e. a theory which one trusts non-linearly and not just in perturbation theory. This is why in our fundamental theory of gravity the $+++$ amplitude must be zero.  

A similar argument shows that the amplitudes that we discarded as unphysical in the previous subsection, e.g. the $+--$ amplitude $M_p^3 \ket{2}{3}^{-6}\ket{1}{2}^2\ket{1}{3}^2$, cannot come from any local cubic interaction vertex, and so are of no physical interest. 

All in all, with the assumption that gravitons are described by the metric so that the helicity states are given by (\ref{helicity-h}), and the requirement of not higher than second order field equations, we conclude that only the following two 3-graviton amplitudes may be non-zero
\be
{\cal M}^{+--} = \frac{1}{M_p} \frac{\bra{2}{3}^6}{\bra{1}{2}^2 \bra{1}{3}^2}, \qquad
{\cal M}^{-++} = \frac{1}{M_p} \frac{\ket{2}{3}^6}{\ket{1}{2}^2 \ket{1}{3}^2}.
\ee
Here the couplings in front may in principle be different, but if we in addition assume parity invariance (or unitarity in the case of Lorentzian signature) we get the above expressions. 

Having fixed the form of the 3-graviton amplitudes, one can show that higher tree-level amplitudes are determined by factorisation. Below we will see how this works for the 4-graviton amplitudes. This completes the construction of the tree-level theory of interacting gravitons. Once again, we remind the reader that the only assumptions that entered into the argument was that the gravitons are described by the metric, hence (\ref{helicity-h}) form of the helicity spinors, and that there must not be higher than second derivatives in the field equations. 

\section{Chiral description of gravitons and a new cubic vertex}
\label{sec:conn}

\subsection{Representations of Lorentz group}

It is time to question our assumption that the gravitons must be described by a symmetric rank two tensor. We can justify this assumption by our desire to describe spin 2 particles. In the spinor language the object $h_{ABA'B'}$ has 4 spinor indices, and is thus indeed spin 2, each spinor index contributing spin one half. So, this object is definitely capable of describing spin 2 particles, as we already know. We can also justify the use of $h_{\mu\nu}$ by recourse to Einstein who taught us that gravity is geometry. 

However, there are other objects with 4 spinor indices, e.g. an object of the type $a_{ABCA'}$ or a completely chiral object $\psi_{ABCD}$. They are both spin 2, so could they be used to describe gravitons? And as for the recourse to Einstein, it can be objected that gravity is indeed geometry, but there is more to geometry than metric geometry. 

The real reason why we use $h_{\mu\nu}$ to describe gravity is because there are real objects of the type $h_{ABA'B'}$. Indeed, in Minkowski signature the operation of complex conjugation interchanges the primed and unprimed spinors. Thus, there are objects of the type $h_{ABA'B'}$ that go into themselves under the complex conjugation. These are real metrics.

However, one of the most useful operations in theoretical physics is to allow real quantities to become complex. We have already done such an analytic continuation to when we considered the 3-graviton amplitudes above, as these required analytic continuation to complex momenta to be non-vanishing. So, why don't we try to use intrinsically complex objects such as $a_{ABCA'}$ or $\psi_{ABCD}$ to describe gravitons? In fact, it is precisely the object $\psi_{ABCD}$ that is used to describe gravitons of one of the helicities in the non-linear graviton construction \cite{Penrose:1976js} of Penrose. So, we know that it is possible to work with objects other than the metric, and in particular with complex objects. Of course, at some point we will need to understand what "real" gravitons correspond to, i.e. how reality conditions can be imposed. But we can postpone this question till we see if the game is worth playing. 

So, the idea is to change the representation of the Lorentz group used to describe gravitons, and instead of a parity-even representation $S_+^2\otimes S_-^2$, try to use asymmetric representations $S_+^3\otimes S_-$. Here $S_\pm^k$ stand for the symmetric tensor product of the corresponding spinor representations, with ${\rm dim}(S_\pm^k)=k+1$. The other possibility, namely $S_+^4$ seems to be well-suited for describing gravitons of just a single helicity, but not both. So, we will not consider it in any details in this paper, see, however, some further comments in the last section. 

\subsection{Chiral gravitons}

Our starting point will be that it is possible to describe gravitons, i.e. massless spin two particles (of both helicities) by a field $a_{ABCA'}\in S_+^3\otimes S_-$. We do not need to state the corresponding field equation here, as we did not start with the linearised Einstein equations when we treated the metric case. It suffices to know that such description of gravitons is possible. A very explicit discussion, including a treatment of the reality conditions and the mode decomposition can be found in \cite{Delfino:2012zy}. A description of the associated complex of differential operators can be found in \cite{Krasnov:2014wha}. But the only thing that we need the reader to accept for purposes of this paper is that it is possible to use the representation $S_+^3\otimes S_-$ to describe massless spin two particles.

The next thing that we need is an expression for the helicity spinors. Again, these can be derived from a careful Hamiltonian analysis of what the physical states are. Such a derivation is presented in \cite{Delfino:2012aj}. But the only thing we need for purposes of this paper is the final expression. It is build in exact analogy to what the helicity spinors are in the metric case, so it could have been guessed even prior to any analysis
\be\label{helicity-a}
\epsilon^-_{ABCA'} = M \frac{q_A q_B q_C k_{A'}}{\ket{q}{k}^3}, \qquad
\epsilon^+_{ABCA'} = \frac{1}{M} \frac{k_A k_B k_C q_{A'}}{\bra{k}{q}}.
\ee
The choice of the spinor numerators is more or less dictated by the analogy with the metric case. The denominators are chosen so that the helicity spinors are homogeneity degree zero functions of the reference spinors $q_A, q_{A'}$. Note also that the scaling properties of the helicity spinors are as required. However, and this is the central point, because we are now using a chiral representation of the Lorentz group the mass dimensions of the arising spinors are different from zero, as in the metric case. This has to be corrected by an introduction of dimensionful parameters $M, 1/M$ in (\ref{helicity-a}). It is the appearance of these parameters that will change the dimension count and will allow for more interactions to be possible. 

Because of the central role played by the factors of $M, 1/M$ we need to discuss the appearance of the scale $M$ further. Here we have seen the necessity of $M$ just from a dimensional point of view, to get the right dimensionless helicity spinors. The need for the factors, as well as the fact that it is the same scale that appears in both expressions, can also be understood by analysing the (Lorentzian signature) reality conditions for the gravitons in the $S_+^3\otimes S_-$ description. The basis idea of this reality conditions is that, while the complex conjugate object $(a^*)_{A'B'C'A}\in S_-^3\otimes S_+$ cannot be equated to the original object in $S_+^3\otimes S_-$, it is possible to apply to $a_{ABCA'}$ the Dirac operator twice, and obtain an object in  $S_+^3\otimes S_-$. Indeed, recall that the Dirac operator acts by flipping a spinor into a spinor of opposite helicity $\partial\!\!\!\slash: S_+\to S_-$. Thus, we get $\partial\!\!\!\slash^2 a\in S_-^3\otimes S_+$, and therefore we can equate 
\be\label{reality}
\frac{1}{M^2} \partial\!\!\!\slash^2 a = a^*.
\ee
It turns out that these are the correct reality conditions to be used, in the sense that they in particular guarantee that the metric constructed as $h=(1/M) \partial\!\!\!\slash a$ is real. Once the reality conditions are imposed one derives (\ref{helicity-a}) as the correct expressions for the spinor helicities. We refer the reader to \cite{Delfino:2012zy} for more details. 

Even though the reality condition (\ref{reality}) provide a relation between the two helicities, and require that it is the same scale M that is used in both, one does not impose any reality conditions in the complexified setting, and so one cannot refer to (\ref{reality}). In this case the only justification for the presence of $M, 1/M$ in (\ref{helicity-a}) is from the dimensions count. There is then no reason why for the numerical coefficients in front of the two helicities to be the same. 

The final point is about the meaning of the scale $M$. We see that it is only possible to describe gravitons using $a\in S_+^3\otimes S_-$ if one introduces a scale $M$. This scale turns out to be related to the radius of curvature of the background. In other words, this description of gravitons is only possible on spaces of non-zero scalar curvature, e.g. $AdS_4$ or $dS_4$, with $M$ being related to the cosmological constant $M^2=\pm \Lambda/3$. Thus, the scale $M$ is just the inverse of the radius of curvature of the background on which we describe gravitons, and the $S_+^3\otimes S_-$ only works for $\Lambda\not =0$. 

Our final remark in this subsection is about the meaning of the helicity spinors (and Fourier transform and null momentum vectors) on a constant curvature space. This can be explained by making an assumption that we are interested in gravitons with energies much larger than $M$. For these gravitons we can certainly neglect the effects of living on a non-flat space, and this justifies the usage of the Fourier transform. We cannot, however, completely forget about the scale $M$ because it is necessary for dimensional reasons. So, some of the formulas below, while pertaining to sufficiently energetic gravitons living in effective flat space, will contain factors of $M$. For a further justification of such rules the reader is directed to \cite{Delfino:2012aj}. 

\subsection{Dimensional analysis}

Having expressions (\ref{helicity-a}) for the helicity spinors in our hands, we can repeat the same dimensional analysis as was done in the case of the metric gravity, and determine which amplitudes can come from a Lagrangian without higher derivatives. The central point here is that because of the presence of dimensionful factors in (\ref{helicity-a}) the dimensional count changes completely, and was not possible in the case of metric gravity becomes possible here.

Let us first consider the $--+$ amplitude given by ${\cal M}^{--+}=(1/M_p) \bra{1}{2}^6/\bra{1}{3}^2\bra{2}{3}^2$. To get this amplitude we must take some cubic interaction term in the Lagrangian and replace in it the 3 occurrences of $a$ by two negative helicity spinors and one positive. The resulting factors of $M$ are $M^2 (1/M)=M$. This means that there must be a factor of $1/M$ in the cubic interaction term so that the result is as desired. All in all, we see that this amplitude is produced by a cubic interaction of the type
\be\label{L1}
{\cal L} \sim \frac{1}{MM_p} (\partial a)^3.
\ee
Importantly, we need one more derivative in this cubic vertex as compared to the metric case, because we need an additional factor of $1/M$ to cancel what comes from the helicity states. Note that in spite of the presence of an additional derivative here, this cubic interaction does not lead to higher order field equations. It simply leads to non-linearity in the first derivative. So, this is an acceptable interaction. 

Let us now analyse the all plus amplitude ${\cal M}^{+++}=\alpha M_p^{-5} \ket{1}{2}^2\ket{1}{3}^2\ket{2}{3}^2$. We should get this amplitude for a cubic interaction term in the Lagrangian by replacing 3 occurrences of $a$ by 3 positive helicity spinors. This gives a factor of $1/M^3$ that should not appear in the final answer, and thus has to be cancelled by a factor of $M^3$ in the interaction. Thus, we see that the Lagrangian that would produce this amplitude is
\be\label{L2}
{\cal L} \sim \frac{\alpha M^3}{M_p^5} (\partial a)^3,
\ee
which is, apart from a different pre factor, schematically the same as (\ref{L1})! Unlike in the metric case, where to get this amplitude one requires an interaction with as many as 6 derivatives, in the $S_+^3\otimes S_-$ description we need exactly the same number of derivatives as in the interaction that leads to the usual $--+$ amplitude. It is clear that this happens because of the presence of factors of $M$ in the helicity spinors. We note that of course the index contractions in (\ref{L1}) and (\ref{L2}) may be different, but what matters for our discussion here is just the total number of derivatives, and in this respect both interactions behave in the same way. 

The conclusion now is that there is no reason why the amplitude ${\cal M}^{+++}$ must be set to zero, as it is in absolutely no conflict with the desired second order nature of the field equations. So, as claimed in the Introduction, in the $S_+^3\otimes S_-$ description it is possible to add a new cubic interaction (and a new coupling) that leads to the non-zero ${\cal M}^{+++}$ amplitude. This interaction is of schematically the same form as the interaction (\ref{L1}) that leads to the amplitudes familiar from GR.

It remains to consider the other two helicity configurations. Let us start with $---$ which is proportional to $M_p^{-5} \bra{1}{2}^2\bra{1}{3}^2\bra{2}{3}^2$. We would like to get this amplitude by replacing 3 copies of $a$ in the cubic vertex with 3 copies of $\epsilon^-$, each carrying a factor of $M$. Thus, there must be a factor of $M^{-3}$ in the vertex. It is immediately clear that there must be as many as 9 derivatives in such a vertex, which is too many to have second order field equations. So, we cannot have a non-zero ${\cal M}^{---}$ amplitude. Therefore, if we insist on parity-invariance, then we should also set to zero the all plus amplitude. But in the absence of such a requirement, we learn that the $+++$ amplitude is possible while $---$ is not.

The analysis for the helicity configuration $-++$ is a bit more tricky, because it appears that there is now a factor of $1/M$ that comes from the helicity spinors that must be cancelled by a factor of $M$ in the vertex. Thus, it would seem that the required interaction is of the form
\be
{\cal L}\sim \frac{M}{M_p} a^2 (\partial a),
\ee
which is of the Yang-Mills type. This is essentially correct, with the only subtlety being that the vertex (\ref{L1}) also gives a contribution to the same amplitude. This comes from the fact that the positive helicity gravitons should be kept slightly massive (with a mass of $M^2$), and the mass is to be set to zero only at the end of the calculation. See \cite{Delfino:2012aj} for a discussion on this rule, as well as the determination of the $-++$ amplitude. 

All in all, we learn that in the $S_+^3\otimes S_-$ formalism it is possible to have the usual $-++$ and $--+$ amplitudes, as well as the new amplitude $+++$ that was previously impossible in the metric formalism. The amplitude $---$ continues to be forbidden. 

We have so far worked in the complexified setting. If one wants the theory to describe Lorentzian signature gravitons, one has to impose the reality conditions (\ref{reality}). If one then wants the arising S-matrix to be unitary, one needs to require parity symmetry. Since the all minus amplitude is forbidden by the requirement of the absence of higher derivatives, this also forbids the all plus amplitude. Thus, the imposition of the reality conditions (\ref{reality}) as well as the requirement of unitarity brings us back to the familiar GR amplitudes. But we see that there is more freedom in the complexified setting, and other graviton interactions are possible in the $S_+^3\otimes S_-$ description, while not leading to higher derivatives. 

\section{4-graviton amplitudes from factorisation}
\label{sec:unitarity}

In GR the higher tree-level amplitudes can be determined from the requirement that they factorise correctly on their physical singularities. For the 4-graviton amplitude the first argument of this type seems to be that in paper \cite{Grisaru:1975bx}. The more modern version of the same argument uses spinor helicity, see e.g. the talk \cite{Nima} as well as related analysis in \cite{McGady:2013sga}. A recent paper \cite{Zhou:2014yaa} attempts to run the same construction for higher numbers of scattering particles. The factorisation argument generalises directly to the complexified setting, and so applies also to our "S-matrix" of holomorphic gravitons. 

Our task in this section is to determine what the presence of the new $+++$ amplitude implies for the higher particle number amplitudes. Here we discuss the 4-graviton case. While some of the amplitudes are unchanged from the familiar GR expressions, we need to check what is modified and what is not in our theories, so we will work out all amplitudes. We will see that the $----$ and $---+$ amplitudes continue to be zero as in GR, the MHV amplitude $--++$ is unchanged from what it is in GR, while there are now non-vanishing $-+++$ and $++++$ amplitudes. Moreover, in the case of the $++++$ amplitude, apart from a contribution that is fixed by the factorisation, one can add a new non-singular in $s,t,u$ variables amplitude with a new coupling constant, and this amplitude arises from an interaction that does not lead to higher derivatives. 

\subsection{The $----$ and $---+$ amplitudes}

It is not hard to see that these must continue to be zero as in GR. Indeed, by factorisation, in the limit when e.g. $k_{12}=k_1+k_2$ becomes null $s=k_1\cdot k_2\to 0$, the amplitude must behave as $1/s$ times the product of two 3-graviton on-shell amplitudes. This can either be a helicity assignment on the internal line as shown in figure below
$$\parbox{25mm}
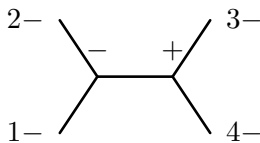
{\begin{fmfgraph*}(25,15) 
 \fmfleft{i1,i2} 
 \fmfright{o1,o2}
 \fmfv{label=$1-$,label.angle=180}{i1}
 \fmfv{label=$4-$,label.angle=0}{o1}
 \fmfv{label=$2-$,label.angle=180}{i2}
 \fmfv{label=$3-$,label.angle=0}{o2}
 \fmfv{label=$-$,label.angle=90}{v1}
 \fmfv{label=$+$,label.angle=90}{v2}
\fmf{plain}{i1,v1,i2} 
\fmf{plain}{o1,v2,o2} 
\fmf{plain}{v1,v2}
\end{fmfgraph*}}$$
or an opposite assignment where the $12i$ vertex, $i$ being the internal momentum, is $--+$ and the $34i$ vertex is $---$. For either of these, the $---$ amplitude is equal to zero, hence there cannot be a contribution to $----$ amplitude that has singularities in the $s,t,u$ variables. Of course, it is always possible to have non-singular contributions, but it is an easy dimension count exercise to see that these cannot result from any vertex without higher derivatives. 

The $---+$ amplitude is more subtle, as it now seems to be possible to have a helicity assignment on the internal line that does not obviously lead to a zero answer
$$\parbox{25mm}
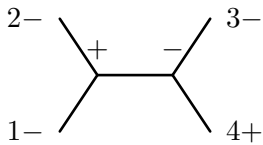
{\begin{fmfgraph*}(25,15) 
 \fmfleft{i1,i2} 
 \fmfright{o1,o2}
 \fmfv{label=$1-$,label.angle=180}{i1}
 \fmfv{label=$4+$,label.angle=0}{o1}
 \fmfv{label=$2-$,label.angle=180}{i2}
 \fmfv{label=$3-$,label.angle=0}{o2}
 \fmfv{label=$+$,label.angle=90}{v1}
 \fmfv{label=$-$,label.angle=90}{v2}
\fmf{plain}{i1,v1,i2} 
\fmf{plain}{o1,v2,o2} 
\fmf{plain}{v1,v2}
\end{fmfgraph*}} $$
This must behave as $1/s$ times the product of two amplitudes each of which appears to be non-zero. But note that only the amplitudes $--+$ that are present in GR appear here. We then know that $---+$ amplitude continues to be zero in GR, and so we don't need to analyse this case further. 

\subsection{The $--++$ amplitude}

This is the only 2-2 graviton amplitude that is non-zero in GR. The only contribution to the $s$-channel singularity is
$$\parbox{25mm}
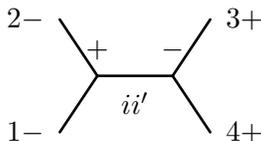
{\begin{fmfgraph*}(25,15) 
 \fmfleft{i1,i2} 
 \fmfright{o1,o2}
 \fmfv{label=$1-$,label.angle=180}{i1}
 \fmfv{label=$4+$,label.angle=0}{o1}
 \fmfv{label=$2-$,label.angle=180}{i2}
 \fmfv{label=$3+$,label.angle=0}{o2}
 \fmfv{label=$+$,label.angle=90}{v1}
 \fmfv{label=$-$,label.angle=90}{v2}
\fmf{plain}{i1,v1,i2} 
\fmf{plain}{o1,v2,o2} 
\fmf{plain,label=$ii'$}{v1,v2}
\end{fmfgraph*}} $$
The other assignment of helicities on the internal line is not possible because there is no $---$ amplitude. So, we see that the contribution from this channel is unchanged from what it is in GR. Let us work out what this $s\to 0$ behaviour must be. We have the product of two $--+$ and $-++$ amplitudes
\be
\frac{1}{s} \frac{1}{M_p^2}  \frac{\bra{1}{2}^6}{\bra{1}{i}^2\bra{2}{i}^2} \frac{\ket{3}{4}^6}{\ket{i}{3}^2\ket{i}{4}^2}.
\ee
We can remove all occurrences of $i,i'$ from this formula using the momentum conservation $11'+22'=ii'=33'+44'$. From this we get $\ket{3}{i} \bra{1}{i}= \ket{3}{2}\bra{1}{2}=\ket{3}{4}\bra{1}{4}$ and $\ket{4}{i}\bra{2}{i} = \ket{4}{1}\bra{2}{1}=\ket{4}{3}\bra{2}{3}$, and thus the $s$-channel behaviour is
\be
\frac{1}{s} \frac{1}{M_p^2}  \bra{1}{2}^4 \ket{3}{4}^4 \frac{1}{\ket{2}{3}^2\bra{2}{3}^2} =
\frac{1}{s} \frac{1}{M_p^2}  \bra{1}{2}^4 \ket{3}{4}^4 \frac{1}{t^2},
\ee
where $t=\ket{2}{3}\bra{2}{3}$. 

Let us now consider the $t$-channel singularity. The possible helicity assignments on the internal line are
$$\parbox{45mm}
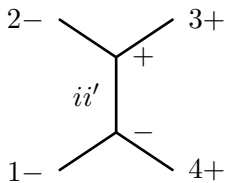
{\begin{fmfgraph*}(15,25) 
 \fmftop{i1,i2} 
 \fmfbottom{o1,o2}
 \fmfv{label=$2-$,label.angle=180}{i1}
 \fmfv{label=$1-$,label.angle=190}{o1}
 \fmfv{label=$3+$,label.angle=0}{i2}
 \fmfv{label=$4+$,label.angle=0}{o2}
 \fmfv{label=$+$,label.angle=0}{v1}
 \fmfv{label=$-$,label.angle=0}{v2}
\fmf{plain}{i1,v1,i2} 
\fmf{plain}{o1,v2,o2} 
\fmf{plain,label=$ii'$}{v1,v2}
\end{fmfgraph*}} $$
and the opposite assignment. However, because in the on-shell cubic vertex it is either angle bracket or square bracket contractions that are non-zero, we see that only one of these two possible assignments gives a non-zero result. Let us assume that this is the assignment shown. It gives the following $t\to 0$ behaviour of the amplitude
\be
 \frac{1}{t} \frac{1}{M_p^2}  \frac{\ket{3}{i}^6}{\ket{2}{3}^2\ket{2}{i}^2} \frac{\bra{1}{i}^6}{\bra{1}{4}^2\bra{i}{4}^2} .
\ee
Using the momentum conservation $22'+33'=ii'=11'+44'$ we get $\ket{2}{i}\bra{4}{i}=\ket{2}{3}\bra{4}{3}=\ket{2}{1}\bra{4}{1}$ and $\ket{3}{i}\bra{1}{i}=\ket{3}{2}\bra{1}{2}=\ket{3}{4}\bra{1}{4}$. Thus, the $t$-channel behaviour is
\be
 \frac{1}{t} \frac{1}{M_p^2}  \bra{1}{2}^4 \ket{3}{4}^4 \frac{1}{\ket{3}{4}^2\bra{3}{4}^2} =
\frac{1}{t} \frac{1}{M_p^2}  \bra{1}{2}^4 \ket{3}{4}^4 \frac{1}{s^2},
 \ee
 where we have used momentum conservation in the form $\bra{1}{2}^2/\bra{1}{4}^2=\ket{3}{4}^2/\ket{2}{3}^2$ to get the first expression. 

The analysis of the $u=\ket{1}{3}\bra{1}{3}$ behaviour is similar. We can summarise the results by saying that the amplitude has the following form
\be
{\cal M}^{--++} = \frac{1}{M_p^2}  \bra{1}{2}^4 \ket{3}{4}^4 F(s,t,u),
\ee
where the function $F(s,t,u)$ has the following behaviour near its singularities
\be
F(s,t,u) = 
 \left\{ \lower5ex\vbox{\hbox{$\frac{1}{s}\frac{1}{t^2},\qquad s\to 0$}\hbox{}\hbox{$\frac{1}{t}\frac{1}{s^2},\qquad t\to 0$}\hbox{}\hbox{$\frac{1}{u}\frac{1}{s^2},\qquad u\to 0$}} \right.
\ee
The only such function is 
\be\label{F}
F(s,t,u)=-\frac{1}{stu},
\ee
which gives the known correct answer for this amplitude. 

\subsection{The $-+++$ amplitude}

The singularities of this amplitude have two possible contributions. In one of them, there are only those helicity assignments that lead to the product of two $++-$ amplitudes present in GR. These cannot give rise to any non-zero answer, and so don't need to be analysed. Thus, the only possible contribution to e.g. the $s$-channel is
$$\parbox{25mm}
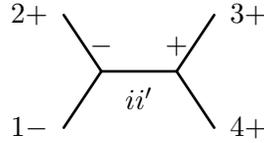
{\begin{fmfgraph*}(25,15) 
 \fmfleft{i1,i2} 
 \fmfright{o1,o2}
 \fmfv{label=$1-$,label.angle=180}{i1}
 \fmfv{label=$4+$,label.angle=0}{o1}
 \fmfv{label=$2+$,label.angle=180}{i2}
 \fmfv{label=$3+$,label.angle=0}{o2}
 \fmfv{label=$-$,label.angle=90}{v1}
 \fmfv{label=$+$,label.angle=90}{v2}
\fmf{plain}{i1,v1,i2} 
\fmf{plain}{o1,v2,o2} 
\fmf{plain,label=$ii'$}{v1,v2}
\end{fmfgraph*}} $$
This contains the new $+++$ amplitude. Thus, as $s\to 0$ the amplitude must behave as
\be
\frac{1}{s} \frac{\alpha}{M_p^6} \frac{\bra{1}{i}^6}{\bra{1}{2}^2\bra{i}{2}^2} \ket{3}{4}^2\ket{i}{3}^2\ket{i}{4}^2.
\ee
We can then use the momentum conservation in the form $\bra{1}{i}^2/\bra{2}{i}^2=\bra{1}{3}^2/\bra{2}{3}^2$ and $\ket{3}{i}\bra{1}{i} = \ket{3}{2}\bra{1}{2}, \ket{4}{i}\bra{1}{i} = \ket{4}{2}\bra{1}{2}$ to write the above as
\be
\frac{1}{s} \frac{\alpha}{M_p^6} \ket{2}{4}^3 \ket{3}{4}^3 \ket{2}{3}^2 \bra{1}{2}\bra{1}{3} \bra{1}{4}^2 \frac{1}{t^2}.
\ee
The pattern is thus clear: Once again we get that the amplitude behaves as
\be
{\cal M}^{-+++} = \frac{\alpha}{M_p^6} \ket{2}{4}^3 \ket{3}{4}^3 \ket{2}{3}^2 \bra{1}{2}\bra{1}{3} \bra{1}{4}^2  F(s,t,u),
\ee
where $F(s,t,u)$ is the function with the same singular behaviour as in the $--++$ case. It is thus the same function (\ref{F}). 

\subsection{The $++++$ amplitude}

Each channel in this case has two possible helicity assignments on the internal line: The assignment shown in the figure
$$\parbox{25mm}
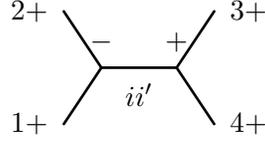
{\begin{fmfgraph*}(25,15) 
 \fmfleft{i1,i2} 
 \fmfright{o1,o2}
 \fmfv{label=$1+$,label.angle=180}{i1}
 \fmfv{label=$4+$,label.angle=0}{o1}
 \fmfv{label=$2+$,label.angle=180}{i2}
 \fmfv{label=$3+$,label.angle=0}{o2}
 \fmfv{label=$-$,label.angle=90}{v1}
 \fmfv{label=$+$,label.angle=90}{v2}
\fmf{plain}{i1,v1,i2} 
\fmf{plain}{o1,v2,o2} 
\fmf{plain,label=$ii'$}{v1,v2}
\end{fmfgraph*}} $$
as well as the opposite assignment. For the assignment shown the singular behaviour must be
\be
\frac{1}{s} \frac{\alpha}{M_p^6} \frac{\ket{1}{2}^6}{\ket{1}{i}^2\ket{2}{i}^2} \ket{3}{4}^2\ket{i}{3}^2\ket{i}{4}^2.
\ee
We now use the momentum conservation to replace $\ket{3}{i}^2/\ket{1}{i}^2 = \ket{2}{3}^2/\ket{1}{2}^2$ and $\ket{4}{i}^2/\ket{2}{i}^2=\ket{1}{4}^2/\ket{1}{2}^2$. We get
\be\label{4+}
\frac{1}{s} \frac{\alpha}{M_p^6} \ket{1}{2}^2 \ket{2}{3}^2 \ket{1}{4}^2 \ket{3}{4}^2 =
\frac{1}{s} \frac{\alpha}{M_p^6} \bra{1}{3}^2 \ket{1}{3}^2 \ket{1}{2}^2 \ket{2}{3}^2 \ket{1}{4}^2 \ket{3}{4}^2 \frac{1}{u^2},
\ee
where in the final expression we wrote the answer in a suggestive form. 

The opposite assignment of helicities in the same channel gives
\be
\frac{1}{s} \frac{\alpha}{M_p^6} \ket{1}{2}^2\ket{i}{1}^2\ket{i}{2}^2 \frac{\ket{3}{4}^6}{\ket{3}{i}^2\ket{4}{i}^2}. 
\ee
Using the momentum conservation $\ket{1}{i}^2/\ket{3}{i}^2 = \ket{1}{4}^2/\ket{3}{4}^2, \ket{2}{i}^2/\ket{4}{i}^2=\ket{2}{3}^2/\ket{3}{4}^2$ this reduces to the same expression (\ref{4+}). This means that the total contribution from the $s$-channel is a multiple of (\ref{4+}). In this paper we do not keep track of numerical factors, and so we simply take (\ref{4+}) as the $s$-channel behaviour. 

Similarly, the $t$-channel gives
\be
\frac{1}{t} \frac{\alpha}{M_p^6} \ket{1}{3}^2 \ket{2}{3}^2 \ket{2}{4}^2 \ket{1}{4}^2 =
\frac{1}{t} \frac{\alpha}{M_p^6} \bra{1}{3}^2 \ket{1}{3}^2 \ket{1}{2}^2 \ket{2}{3}^2 \ket{1}{4}^2 \ket{3}{4}^2 \frac{1}{s^2},
\ee
where we used the momentum conservation to get the last expression. 

Overall, we see that the part of the amplitude that is determined by the factorisation is given by
\be\label{all-plus}
{\cal M}^{++++} = \frac{\alpha}{M_p^6} \bra{1}{3}^2 \ket{1}{3}^2 \ket{1}{2}^2 \ket{2}{3}^2 \ket{1}{4}^2 \ket{3}{4}^2 F(s,t,u),
\ee
where again the same function (\ref{F}) makes appearance. 

\subsection{The non-singular part of the $++++$ amplitude}

It is easy to see that in addition to (\ref{all-plus}) there can also be an $s,t,u$-non-singular contribution to the all plus amplitude. Indeed, we can add
\be
{\cal M}^{++++}_{\rm new} = \frac{\beta}{M^8_p} \left(\ket{1}{3}^4 \ket{2}{4}^4 + \ket{1}{2}^4 \ket{3}{4}^4 +\ket{1}{4}^4 \ket{2}{3}^4 \right),
\ee
where $\beta$ is a new coupling constant. The expression in brackets is completely symmetric, and has the right little group transformation properties. It has no singularities as $s,t,u\to 0$, and so it cannot be determined from the factorisation. Usually this possible contribution to the amplitude is discarded because it can only come from an interaction that leads to higher derivatives. However, with our helicity spinors the required interaction is of the form
\be
\frac{\beta M^4}{M_p^8} (\partial a)^4,
\ee
and is allowed. We see that this is just the beginning of a general pattern that at each graviton order there is a new possible all plus amplitude, with a new coupling constants (or several new coupling constants). All this is possible with keeping just second derivative nature of the field equations of the theory. 

\section{Higher amplitudes and BCFW recursion}
\label{sec:BCFW}

In principle, it should be possible to compute also the higher graviton number amplitudes using factorisation, see e.g. \cite{Zhou:2014yaa} for an attempt in this direction. However, in practice, usage of the BCFW recursion relation is much more powerful. In this section we explain that some amplitudes continue to be computable using the BCFW recursion even for our infinite-parametric class of theories. The pattern below is exactly analogous to what is at play for a certain infinite-parametric family of gauge-theories, see \cite{Cofano:2015jva}.

\subsection{The $-+++$ amplitude via BCFW}

Among the already encountered amplitudes, the MHV amplitude $--++$ is the same as in GR, and is thus clearly computable using the BCFW recursion. The new $+++$ 3-graviton amplitude cannot contribute to this recursion because it would need to be taken together with $---$ that is zero. This is the BCFW explanation why the amplitude is unchanged from its GR value. 

To see if we can apply BCFW to other amplitudes, let us rewrite the $-+++$ amplitude replacing $F(s,t,u)$ by its expression in terms of spinor contractions
\be\label{a-3p}
{\cal M}^{-+++} = \frac{\alpha}{M_p^6} \bra{1}{4} \frac{\ket{2}{4}^3\ket{3}{4}^3\ket{2}{3}^2}{\ket{1}{2}\ket{1}{3}\ket{1}{4}}.
\ee
Written in this way, the leg $4$ is made special. However, using the momentum conservation it is easy to see that the amplitude is in fact symmetric in legs $2,3,4$. 

Let us now consider the following BCFW shift
\be
1\to 1+ z 4, \qquad 4'\to 4'-z1'.
\ee
It is clear that only the angle brackets in the denominator are sensitive to the shift. Thus, under this shift the $-+++$ amplitude behaves as $1/z^2$. This is the familiar from GR behaviour. 

Because the amplitude tends to zero as $z\to\infty$, we can determine it using the BCFW. Let us see how this works. There are only two contributing diagrams
$$\parbox{25mm}
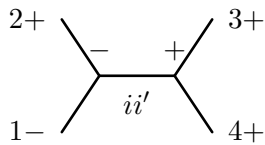
{\begin{fmfgraph*}(25,15) 
 \fmfleft{i1,i2} 
 \fmfright{o1,o2}
 \fmfv{label=$1-$,label.angle=180}{i1}
 \fmfv{label=$4+$,label.angle=0}{o1}
 \fmfv{label=$2+$,label.angle=180}{i2}
 \fmfv{label=$3+$,label.angle=0}{o2}
 \fmfv{label=$-$,label.angle=90}{v1}
 \fmfv{label=$+$,label.angle=90}{v2}
\fmf{plain}{i1,v1,i2} 
\fmf{plain}{o1,v2,o2} 
\fmf{plain,label=$ii'$}{v1,v2}
\end{fmfgraph*}} $$
and the same diagram with $1$ first connecting to $3$. For the diagram shown the on-shell momentum on the internal line is
\be
(1+z_p 4)1'+22'=2\left( 2'+\frac{\ket{1}{4}}{\ket{2}{4}} 1'\right),
\ee
where $z_p=\ket{1}{2}/\ket{2}{4}$. The contribution to the amplitude from this BCFW pole is thus given by
\be\nonumber
\frac{\alpha}{M_p^6} \frac{1}{\ket{1}{2}\bra{1}{2}} \frac{\bra{1}{i}^6}{\bra{1}{2}^2\bra{i}{2}^2} \ket{3}{4}^2\ket{i}{3}^2\ket{i}{4}^2 =
\frac{\alpha}{M_p^6} \bra{1}{2} \frac{\ket{3}{4}^2\ket{2}{3}^2\ket{2}{4}^4}{\ket{1}{2}\ket{1}{4}^2}=
-\frac{\alpha}{M_p^6} \frac{\ket{2}{4}^3\ket{3}{4}^3\ket{2}{3}^2}{\ket{1}{2}\ket{1}{3}\ket{1}{4}^2}\ket{2}{4}\bra{2}{4}.
\ee
The other contribution is obtained by exchanging $2$ with $3$ in this formula. Adding the two contributions we get
\be
-\frac{\alpha}{M_p^6} \frac{\ket{2}{4}^3\ket{3}{4}^3\ket{2}{3}^2}{\ket{1}{2}\ket{1}{3}\ket{1}{4}^2} \left( \ket{2}{4}\bra{2}{4}+\ket{3}{4}\bra{3}{4}\right)=\frac{\alpha}{M_p^6} \bra{1}{4} \frac{\ket{2}{4}^3\ket{3}{4}^3\ket{2}{3}^2}{\ket{1}{2}\ket{1}{3}\ket{1}{4}},
\ee
which is precisely (\ref{a-3p}). 

\subsection{Higher amplitudes with at least one minus}

We thus see that both $--++$ and $-+++$ amplitudes can be determined using the BCFW recursion. In the latter case, the particular shift that works is for the unprimed spinor of the negative helicity graviton, and the primed spinor of one of the positive helicity gravitons. Note that, unlike with GR amplitudes, we cannot shift the $-+++$ amplitude on a pair of positive helicity legs, e.g. $2$ and $3$. Indeed, under this shift the amplitude goes as $z^2$ as $z\to\infty$, which makes this shift useless from the point of view of determining the amplitude. So, we see that, as far as the $--++$ and $-+++$ amplitudes are concerned, the only shift that works for both of them it is shift on a negative-positive helicity pair. 

We now give a general argument of the type described in \cite{ArkaniHamed:2008yf} to show that any amplitude with at least one negative helicity graviton is determinable by the BCFW recursion. The shift that is to be performed is to shift the unprimed spinor of the negative helicity leg, and a primed spinor of one of the positive helicity gravitons. 

The only property that our argument will require is that the background field expansion of the action generates a kinetic term for the perturbation with at most two derivatives. This directly follows from the second derivative nature of the field equations for our class of theories, but can also be verified explicitly, see e.g. the related discussion in \cite{Groh:2013oaa}. 

The $1/z^2$ behaviour of the amplitudes then follows from a simple counting. Indeed, under the shift described the helicity spinors of the two shifted gravitons behave as follows
\be
\epsilon^- \sim \frac{1}{z^3}, \qquad \epsilon^+ \sim \frac{1}{z}.
\ee
In both cases the $z$-dependence comes from the denominators containing the factors $\ket{q}{k}^3$ and $\bra{k}{q}$. 

We can now estimate the $z$-behaviour of the shifted amplitude by considering a propagating of a 
very energetic graviton on a background of soft gravitons. We are interested in the amplitude for the hard graviton not to change its helicity. This can be computed from the quadratic part of the linearisation of the action of the theory on the background of soft gravitons. We are then to project this linearised action onto the helicity states of the hard graviton. Given the fact that there is a factor of $1/z^4$ coming from the helicity states, and the fact that there are just two derivatives in the linearised Lagrangian that can produce at most a factor of $z^2$, we see that the shifted amplitudes can go at most as $1/z^2$ as $z\to\infty$.

Thus, we see that the shift on a $-+$ pair is always applicable for our class of theories, and amplitudes that have at least one minus can be obtained from the associated BCFW recursion relation. This is like in GR, except that now we have more possible basic amplitudes that can contribute to the recursion, as we have seen in the above example of the $-+++$ amplitude, where the $+++$ amplitude played role. 

The fact that BCFW is still applicable allows for a useful observation: All MHV amplitudes are unchanged from their GR values. By an MHV amplitude we mean an amplitude with just two plus helicity gravitons. This statement is easily proven by induction. Indeed, we know that the $--++$ amplitude is what it is in GR, as are the $----$ and $---+$ plus amplitudes that continue to be zero as in GR. Then the fact that the all minus amplitude for any number of gravitons must be zero can be seen to follow e.g. from factorisation, as such an amplitude would necessarily use a lower order all minus amplitude. The fact that all minus one plus amplitude equals to zero follows from the BCFW by induction. We can then use these facts to determine the all minus two plus MHV amplitudes by recursion. Let us assume that an $n$-graviton MHV amplitude is the same as in GR. Then the BCFW contributions to the $n+1$ MHV amplitude are all of the type $n$-th order MHV amplitude times a $--+$ amplitude. These are the same as in GR, and so the $(n+1)$-th MHV amplitude is the same as in GR. 

Another remarks is that, as we noted above, unlike in GR, we can no longer continue on other helicity pairs. Indeed, for the $-+++$ amplitude the $++$ pair shift did not lead to a vanishing at large $z$ behaviour. It is possible that the $--$ pair shift still works for our theories, but this is not of much help because it cannot be used to determine the new mostly plus amplitudes that are non-zero for our theories. Thus, the most powerful shift is that for a $-+$ pair, as it can be used to determine the amplitudes with at least one minus in them.

Let us see how the program of determining the graviton amplitudes using the BCFW can work in practice. We know that at graviton order $n$ we can determine all amplitudes with at least one minus. For this we will need to have all amplitudes at order $n-1$, including the all plus amplitude at order $n-1$. However, we know that this amplitude cannot be determined using the BCFW recursion, at least not for the simple shift on a pair of gravitons as considered so far. So, to determine higher amplitudes we will need some way of also determining the all plus amplitudes. 

\subsection{Higher all plus amplitudes} 

For the all plus amplitudes we know that at every graviton order there are new amplitudes that can be added, with new coupling constants. This can be done without increasing the order of field equations, as the required vertex in the Lagrangian is of the form
\be
\frac{M^n}{M_p^{3n-4}} (\partial a)^n.
\ee
This part of the order $n$ all plus amplitude is new, and is just added to recursion process at every next order. 

However, there is also a contribution to the all plus amplitude that comes from lower amplitudes, as we have seen in the above $++++$ example, where we determined such a contribution from factorisation. It is clear that these contributions cannot be obtained from the simple BCFW shift on a pair of gravitons. Indeed, the $++++$ amplitude (\ref{all-plus}) with the function $F(s,t,u)$ replaced with spinor contractions reads
\be
{\cal M}^{++++}\sim \frac{\alpha}{M_p^6} \frac{\ket{2}{3}}{\bra{1}{4}} \ket{1}{2}\ket{1}{3}\ket{1}{4}\ket{2}{3}\ket{2}{4}\ket{3}{4}.
\ee
Performing the BCFW shift on any pair, e.g. $24$, gives that the amplitude behaves as $z^2$ for large $z$, making the BCFW recursion inapplicable. 

We thus need some other recursive way of determining the singular contributions to the all plus amplitudes. In an analogous setup, but in the context of YM-type gauge theories, it was shown in \cite{Cofano:2015jva} that the Risager's shift \cite{Risager:2005vk} is applicable, and can be used to determine the singular part of the all-plus amplitudes from the lower amplitudes. It thus appears that this is also the case for the gravity theories discussed here, but we will not attempt to prove it, leaving this to future work.

\section{Discussion}

In this paper, using the language of graviton scattering amplitudes we have seen how to construct an infinite-parametric family of interacting theories of massless spin two particles. We worked in the complexified setting, where there is no requirement for the scattering amplitudes to assemble into a unitary S-matrix. The possibility of interactions not present in General Relativity arises if one uses a chiral $S_+^3\otimes S_-$ representation of Lorentz group to describe gravitons. We have seen that at every graviton order one can add new all plus graviton amplitudes, with their respective new coupling constants. This is possible while keeping the second order in derivatives character of field equations. 

We then saw how the amplitudes containing at least one minus can be determined from the lower amplitudes using the BCFW recursion. The parts of the all plus amplitudes that follow from lower amplitudes can in principle be reconstructed using factorisation, as we have seen on the example of the $++++$ amplitudes. Another, likely more practical alternative is to use the BCFW recursion based on the Risager's shift \cite{Risager:2005vk}. This works for the $++++$ amplitude, and this also works for all amplitudes in the related setting of "deformations" of YM theory \cite{Cofano:2015jva}. We will no attempt to prove here that it works for all gravity amplitudes, leaving this to future work. Provided this gap can be filled, the described procedure allows to determine all graviton scattering amplitudes starting from the basic ones. The basic amplitudes are the 3-graviton amplitudes present in GR, as well as new all plus amplitude at every graviton order. 

It is important to emphasise that the used here chiral $S_+^3\otimes S_-$ formalism can be used to describe the usual General Relativity. To this end, one imposes the reality conditions (\ref{reality}) that effectively says that the metric is of the Lorentzian signature and real. The requirement of unitarity of the S-matrix then requires that amplitudes are related to their complex conjugates. At the level of the 3-graviton amplitude, the absence of the $---$ amplitude then requires also the $+++$ amplitude to be zero. Similarly, the new all plus amplitudes possible at every particle order are incompatible with unitarity, and must be set to zero. One obtains the usual graviton interactions present in General Relativity. 

The question is then whether it is possible to find reality conditions such that the new theories described here become unitary. We do not know an answer to this at present, so at the moment the infinite parametric theories of interacting gravitons that are the subject of this paper exist only in the complexified setting. 

This paper has words "GR uniqueness" in the title, so it is prudent to discuss the status of GR uniqueness in light of our results. Because of the problems with unitarity in the Lorentzian setting (or problems of finding the acceptable real slice), the "deformed" gravity theories that we described do not yet provide examples of physically acceptable Lorentzian signature gravity theories. So, GR continues to remain the only unitary theory of interacting massless spin 2 particles. However, the GR uniqueness statement is often interpreted more broadly, as the statement that in four dimensions apart from Einstein-Hilbert there is no other Lagrangian describing interacting gravitons that leads to second order field equations. In the case one uses the metric tensor to describe gravitons this statement is true no matter what the signature is, or whether one considers complexified gravity. The main point of this paper is that this is no longer true if one uses a different field to describe gravitons. It would of course be a much stronger result if we could provide also a real slide on which our theories admit an interpretation as unitary theories of Lorentzian signature gravitons. This is however a difficult problem that likely requires a very different type of thinking that was adopted in this paper. 

The obtained infinite-parametric family of gravity theories can also be described off-shell, in fact this is how these theories first appeared. The most economical formulation appears to be that in terms of diffeomorphism invariant ${\rm SO}(3)$ gauge theories \cite{Krasnov:2011pp,Krasnov:2011up,Krasnov:2012pd,Fine:2013qta}. The reader is directed to these references for a Lagrangian description of this family of theories, and to \cite{Delfino:2012aj} for the determination of the 2-2 graviton amplitudes using the resulting Feynman rules. It is also important to stress that the usual General Relativity can be described using this formalism, see the same references. 

Another remark is that, while from the constructions of this paper it appears that a non-zero scalar curvature (i.e. non-zero cosmological constant) is a prerequisite for having distinct from GR gravity theories, this is not so. The described here family of parity asymmetric gravity theories can also be formulated for $\Lambda=0$. However, in this case one needs to introduce a certain set of auxiliary fields, and the description is not as transparent as we encountered above. We refer the reader to \cite{Krasnov:2009ik}.

Our construction used the chiral $S_+^3\otimes S_-$ representation to describe gravitons. We have already mentioned that it is also possible to describe the completely chiral $S_+^4$ representation, at least to describe gravitons of one of the helicities. It is instructive to attempt to play the same game as with $S_+^3\otimes S_-$ and see which new interactions become possible by using the $S_+^4$ representation. Thus, let us assume that the helicity spinors are given by the following expressions
\be
\epsilon^-_{ABCD}= M^2 \frac{q_A q_B q_C q_D}{\ket{q}{k}^4}, \qquad \epsilon^+_{ABCD} = \frac{1}{M^2} k_A k_B k_C k_D.
\ee
Thus, we now have one extra power of the $M,1/M$ in these spinors. Let us see if we can obtain e.g. the $--+$ amplitude from some cubic interaction term in the Lagrangian. We now need the factor of $M^2$ from the helicity spinors to cancel, and thus the required vertex should have the pre factor of $M^{-2} M_p^{-1}$. It should thus have an overall 4 derivatives, which is too many to have second order field equations. Thus, with this representation it is not possible to have the $--+$ amplitude and keep the second order character of field equations. The case of $---$ amplitude is even worse, as one now needs as many as 8 derivatives in the vertex. 

For the other amplitudes the situation is better. The count for the $-++$ amplitude is as follows. We need the vertex to cancel $M^{-2}$ coming form the helicity spinors. Thus, the vertex must be of the form $M^2 M_p^{-1} \psi^3$, containing no derivatives at all. Similarly, to obtain the $+++$ amplitude one needs to cancel the factor of $M^{-6}$ from the helicity spinors. Therefore, the required vertex is of the form $M^6 M_p^{-5} \psi^3$. Once again, this contains no derivatives. All in all we see that it is only possible to have the $-++$ and $+++$ amplitudes in the $S_+^4$ description. Both come from the vertex $\psi^3$ without any derivatives. Since it is not possible to describe the $--+$ amplitude present in GR, the $S_+^4$ description does not seem to be worth developing, at least not as a way to describe full General Relativity. It may, however, be interesting to develop it as a formalism for the description of self-dual gravity. 

As we have already mentioned, deformations of the type described here for the case of gravity can also be considered for the usual Yang-Mills theory. The new interactions one would add in this case are non-renormalisable, but do not lead to higher derivatives in field equations. These deformations of Yang-Mills theory are described in \cite{Cofano:2015jva}. 

Our final set of remarks is on why the described here family of parity asymmetric gravity theories may be interesting. First, they provide an interesting class of modified gravity theories that continue to propagate just two degrees of freedom, as in GR. They can be studied as classical theories of gravity in their own right, and exhibit some intriguing properties such as e.g. singularity resolution inside the black holes \cite{Krasnov:2007ky}. 

However, the author's interest to these theories comes mainly from their possible quantum gravity applications. Thus, it appears more and more likely that this whole family of gravity theories behaves at one-loop as General Relativity does --- they are one-loop renormalisable. This was recently shown to be true for the case of their YM cousins, see \cite{Krasnov:2015kva}. It thus appears that all one-loop divergences are removable by either field redefinitions or by renormalisation of the field and the coupling constants. If this is the case for the case of gravity theories considered here, this means that as one changes the energy the infinite number of coupling constants flows. It is in principle possible to determine the arising one-loop RG flow, and work on this is in progress. It would be very interesting to study this flow, and in particular determine its fixed points. This may lead to progress on the elusive problem of UV completion of quantum gravity.

\section*{Acknowledgements} The author was supported by an ERC Starting Grant 277570-DIGT. I am grateful to the members of DIGT group --- Johnny Espin, Marco Cofano, Yannick Herfray, Chi-Hao Fu and Carlos Scarinci --- for many stimulating discussions on the topics of this paper.

\end{fmffile}


\begin{thebibliography}{99}

\bibitem{Capovilla:1989ac} 
  R.~Capovilla, T.~Jacobson and J.~Dell,
  ``General Relativity Without the Metric,''
  Phys.\ Rev.\ Lett.\  {\bf 63}, 2325 (1989).
  
\bibitem{Bengtsson:1990qg} 
  I.~Bengtsson,
  ``The Cosmological constants,''
  Phys.\ Lett.\ B {\bf 254}, 55 (1991).
  
\bibitem{Krasnov:2006du} 
  K.~Krasnov,
  ``Renormalizable Non-Metric Quantum Gravity?,''
  hep-th/0611182.
  
\bibitem{Krasnov:2008zz} 
  K.~Krasnov,
  ``On deformations of Ashtekar's constraint algebra,''
  Phys.\ Rev.\ Lett.\  {\bf 100}, 081102 (2008)
  [arXiv:0711.0090 [gr-qc]].
  
\bibitem{ArkaniHamed:2008gz} 
  N.~Arkani-Hamed, F.~Cachazo and J.~Kaplan,
  ``What is the Simplest Quantum Field Theory?,''
  JHEP {\bf 1009}, 016 (2010)
  [arXiv:0808.1446 [hep-th]].
  
\bibitem{Penrose:1976js} 
  R.~Penrose,
  ``Nonlinear Gravitons and Curved Twistor Theory,''
  Gen.\ Rel.\ Grav.\  {\bf 7}, 31 (1976).
  
\bibitem{Delfino:2012zy} 
  G.~Delfino, K.~Krasnov and C.~Scarinci,
  ``Pure Connection Formalism for Gravity: Linearized Theory,''
  arXiv:1205.7045 [hep-th].
  
\bibitem{Krasnov:2014wha} 
  K.~Krasnov,
  ``Gravitons and a complex of differential operators,''
  arXiv:1406.7159 [hep-th].
  
\bibitem{Delfino:2012aj} 
  G.~Delfino, K.~Krasnov and C.~Scarinci,
  ``Pure connection formalism for gravity: Feynman rules and the graviton-graviton scattering,''
  arXiv:1210.6215 [hep-th].
  
\bibitem{Grisaru:1975bx} 
  M.~T.~Grisaru, P.~van Nieuwenhuizen and C.~C.~Wu,
  ``Gravitational Born Amplitudes and Kinematical Constraints,''
  Phys.\ Rev.\ D {\bf 12}, 397 (1975).

\bibitem{Nima}
Nima Arkani-Hamed, talk at "New geometric structures in scattering amplitudes", Oxford, 22-25 September 2014, http://people.maths.ox.ac.uk/lmason/NGSA14/Films/Nima-Arkani-Hamed.mp4

\bibitem{McGady:2013sga} 
  D.~A.~McGady and L.~Rodina,
  ``Higher-spin massless $S$-matrices in four-dimensions,''
  Phys.\ Rev.\ D {\bf 90}, no. 8, 084048 (2014)
  [arXiv:1311.2938 [hep-th]].

\bibitem{Zhou:2014yaa} 
  K.~Zhou and C.~Qiao,
  ``General tree-level amplitudes by factorization limits,''
  arXiv:1410.5042 [hep-th].

\bibitem{Cofano:2015jva} 
  M.~Cofano, C.~H.~Fu and K.~Krasnov,
  ``Deformations of Yang-Mills theory,''
  arXiv:1501.00848 [hep-th].

\bibitem{ArkaniHamed:2008yf} 
  N.~Arkani-Hamed and J.~Kaplan,
  ``On Tree Amplitudes in Gauge Theory and Gravity,''
  JHEP {\bf 0804}, 076 (2008)
  [arXiv:0801.2385 [hep-th]].

\bibitem{Groh:2013oaa} 
  K.~Groh, K.~Krasnov and C.~F.~Steinwachs,
  ``Pure connection gravity at one loop: Instanton background,''
  arXiv:1304.6946 [hep-th].
  
\bibitem{Risager:2005vk} 
  K.~Risager,
  ``A Direct proof of the CSW rules,''
  JHEP {\bf 0512}, 003 (2005)
  [hep-th/0508206].
  
   
\bibitem{Krasnov:2011pp} 
  K.~Krasnov,
  ``Pure Connection Action Principle for General Relativity,''
  Phys.\ Rev.\ Lett.\  {\bf 106}, 251103 (2011)
  [arXiv:1103.4498 [gr-qc]].
  
\bibitem{Krasnov:2011up} 
  K.~Krasnov,
  ``Gravity as a diffeomorphism invariant gauge theory,''
  Phys.\ Rev.\ D {\bf 84}, 024034 (2011)
  [arXiv:1101.4788 [hep-th]].
  
\bibitem{Krasnov:2012pd} 
  K.~Krasnov,
  ``A Gauge Theoretic Approach to Gravity,''
  Proc.\ Roy.\ Soc.\ Lond.\ A {\bf 468}, 2129 (2012)
  [arXiv:1202.6183 [gr-qc]].

\bibitem{Fine:2013qta} 
  J.~Fine, K.~Krasnov and D.~Panov,
  ``A gauge theoretic approach to Einstein 4-manifolds,''
  arXiv:1312.2831 [math.DG].
  
\bibitem{Krasnov:2009ik} 
  K.~Krasnov,
  ``Effective metric Lagrangians from an underlying theory with two propagating degrees of freedom,''
  Phys.\ Rev.\ D {\bf 81}, 084026 (2010)
  [arXiv:0911.4903 [hep-th]].
  
\bibitem{Krasnov:2007ky} 
  K.~Krasnov and Y.~Shtanov,
  ``Non-Metric Gravity. II. Spherically Symmetric Solution, Missing Mass and Redshifts of Quasars,''
  Class.\ Quant.\ Grav.\  {\bf 25}, 025002 (2008)
  [arXiv:0705.2047 [gr-qc]].
  
\bibitem{Krasnov:2015kva} 
  K.~Krasnov,
  ``One-loop beta-function for an infinite-parameter family of gauge theories,''
  arXiv:1501.00849 [hep-th].
  

\end{thebibliography}
\end{document}